\documentclass[epj]{svjour}

\usepackage{graphics}
\usepackage{amsmath}
\usepackage{amssymb}
\usepackage{epsfig}
\usepackage{graphicx}

\newcommand{\bb}{\begin{equation}}
\newcommand{\en}{\end{equation}}

\begin{document}

\title{Stochastic model for nucleosome sliding in the presence of DNA ligands}

\author{L. Mollazadeh-Beidokhti\inst{1} \and J. Deseigne\inst{2}$^,$\inst{3} \and D. Lacoste \inst{2}
\and F. Mohammad-Rafiee\inst{1} \and H. Schiessel \inst{4}}

\institute{Institute for Advanced Studies in Basic Sciences (IASBS), Zanjan 45195, P.O. Box 45195-1159, Iran
\and Laboratoire de Physico-Chimie Th\'eorique, UMR 7083, ESPCI, 10 rue Vauquelin, 75231 Paris cedex 05, France
\and CEA - Service de Physique de l'Etat Condens\'e, Centre d'Etudes de Saclay, 91191 Gif-sur-Yvette, France
\and Instituut-Lorentz, Universiteit Leiden, Postbus 9506, 2300 RA Leiden, The Netherlands}
\date{\today}
%
\abstract{
Heat-induced mobility of nucleosomes along DNA is an
experimentally well-studied phenomenon. A recent experiment shows
that the repositioning is modified in the presence of minor-groove
binding DNA ligands. We present here a stochastic three-state
model for the diffusion of a nucleosome along DNA in the presence
of such ligands. It allows us to describe the dynamics and the
steady state of such a motion analytically. The analytical results
are in excellent agreement with numerical simulations of this
stochastic process.With this model, we study the response of a nucleosome to an
external force and how it is affected by the presence of ligands.
\PACS{
      {87.15.A-}{Theory, modelling, and computer simulation}   \and
      {87.15.H-}{Dynamics of biomolecules} \and
      {87.15.Vv}{Diffusion}
     }
}

\maketitle

Eucaryotic DNA is packaged inside the nucleus by being wrapped
onto millions of protein cylinders. Each cylinder is an octamer of
eight histone proteins and is associated with a 147 basepairs (bp)
long stretch of DNA \cite{Luger-97}. The resulting complexes, the
nucleosomes, are connected via stretches of linker DNA; typical
linker lengths range from 12 to 70 bp \cite{Widom-92}. From the
crystal structure \cite{Luger-97} one knows that the DNA is bound
to the octamer at 14 binding sites that correspond to the places
where the minor groove of the DNA faces the octamers defining the
binding path to be a left-handed superhelix of one and three
quarter turns.

With around three quarters of its DNA being tightly bound to the
octamers the DNA binding proteins face the challenge that their
target sites might be masked if they happen to be occupied by a
nucleosome. One knows of two possible pathways of how such wrapped
DNA portions become accessible -- at least temporarily. One
possibility is the spontaneous unwrapping of nucleosomal DNA from
the octamer \cite{Polach-95,Li-05,Gerland-06} that provides DNA
binding proteins a window of opportunity to bind to its target.
Another possibility is that the octamer moves as a whole along the
DNA
\cite{Beard-78,Pennings-91,Flaus-98,Flaus-03,Schiessel-rev03,Marko-07},
thereby releasing previously wrapped portions. This so-called
nucleosome sliding is what we study in the current paper.

A typical experimental system to investigate nucleosome
repositioning consists of a DNA template slightly longer than the
wrapped portion (e.g. 207 bp in Ref. \cite{Pennings-91}) that is
complexed with one octamer. The nucleosome position can be
inferred from the electrophoretic mobility of the complex in a
gel. It is found that nucleosome sliding is a slow process and
that it takes a nucleosome around an hour to reposition completely
on such a short DNA fragment. Another important observation is
that the new positions are all multiple of 10 bp (the DNA helical
repeat) apart from the starting position.

Concerning nucleosome sliding there are currently two possible
mechanisms discussed that could explain those experiments
\cite{Flaus-03,Schiessel-rev03}. Both mechanisms have in common
that they rely on defects that are thermally injected into the
wrapped DNA and that traverse the nucleosome thereby causing its
displacement. The reason for assuming defects as the cause of
repositioning rather than the sliding of the octamer as a whole is
that the latter mechanism would require the simultaneous
detachment of all 14 binding sites which is too costly (around 85
$k_B T$). The two kind of defects are 10 bp loop defects
\cite{Schiessel-01,Kulic-03b} and 1 bp twist defects
\cite{Kulic-03,Farshid-04}. A 10 bp loop defect is a bulge that
carries an extra length of 10 bp causing redistribution events of
that step length. These preserve the rotational orientation of the
nucleosome. This fact as well as the predicted value for the
mobility seems to agree with experiments \cite{Kulic-03b}.

The second class of defects, the 1 bp twist defects, carry a
missing or an extra bp. To accommodate such a defect between two
nucleosomal binding sites the DNA needs to be stretched (or
compressed) and twisted (hence the name). A nucleosome mobilized
by twist defects moves via 1 bp jumps. Since the octamer is always
bound to the minor groove of the DNA, the nucleosome performs a
corkscrew motion around the DNA. Alternatively one can say that
the DNA acts as a molecular corkscrew. Since twist defects are
much cheaper than loop defects ($\approx 9 k_B T$ \cite{Kulic-03}
vs. $\approx 23 k_B T$ \cite{Kulic-03b}), twist defects are
expected to make nucleosomes much more mobile than observed in
experiments. However, repositioning experiments are always done in
the presence of so-called nucleosome positioning experiments that
are now known to be wide-spread in eukaryotic genomes
\cite{Segal-06}. Such positioning sequences (like the sea urchin's
5S rDNA sequence of Ref. \cite{Pennings-91}) make use of the fact
that certain bp sequences induce an anisotropic bendability of the
DNA. A nucleosome diffusing via twist defects feels the resulting
energy landscape since in order to move by 10 bp along the DNA it
needs to perform a full $360 ^\circ$ rotation around the DNA
(unlike a nucleosome that moves via 10 bp bulges). In the end,
both mechanisms are predicted to appear very similar in the
experiments mentioned above \cite{Schiessel-rev03}.

There is an experiment \cite{Gottesfeld-02} that hint at twist
defects being responsible for nucleosome sliding. This experiment
is performed on a 216 bp DNA that contains the sea urchin 5S rDNA
sequence in the presence of minor groove binding pyrrole imidazole
polyamides, synthetic ligands that can be designed to bind to
short specific DNA sequences. It was found that the nucleosome
mobility is dramatically reduced when such ligands are added. This
might reflect the fact that a DNA corkscrew motion is sterically
forbidden once a ligand is bound \cite{Farshid-04}.

Finally, the nucleosomal mobility might also be important when a
transcribing RNA polymerase encounters a nucleosome. Experiments
with short DNA fragments carrying single nucleosomes show that in
such a setup a transcribing RNA polymerase (e.g. bacteriophage
polymerase from T7 \cite{Gottesfeld-02} and SP6
\cite{Studitsky-94}) can transcribe the whole fragment, even
though it is partially occupied by a nucleosome. An interpretation
of how the polymerase negotiates with the nucleosome is tricky
since there are at least two possible explanations. The simpler
explanation is that the polymerase crosses the nucleosome in a
loop \cite{Studitsky-94,Bednar-99}. However, the alternative
explanation \cite{Farshid-04} is that the polymerase pushes the
nucleosome in front of it, pushing it off the template; before the
octamer falls off it rebinds at the other DNA terminus.
Interestingly, in the presence of ligands the polymerase stalls
\cite{Gottesfeld-02}, pointing towards the second mechanism. Note
that the former mechanism would also allow a polymerase to
negotiate with an array of nucleosomes, the second does not
(``traffic jam''). In fact, experiments \cite{Heggeler-95} show
that RNA polymerase can only transcribe through an array and leave
it intact if a nuclear cell extract is present, otherwise the
nucleosomes are stripped off the DNA.

In this paper, we study the influence of ligands on the mobility
of a nucleosome along a long sequence of DNA that is either
assumed to be uniform (``random DNA'') or periodic (``nucleosome
positioning sequence''). First we solve a three-state model
inspired from Ref.~\cite{Farshid-04} using a method described in
\cite{David-07,David-08}. There are some data extracted from
experiments in which the mobility of the nucleosome was probed
using electrophoresis methods \cite{Gottesfeld-02}. The results of
the model are then compared with these existing data for short
DNA. Finally, we study the effect of an external force on the
sliding of the nucleosome in this model. In a very simplified way
this force can represent the action of an RNA polymerase or other
DNA binding proteins on a nucleosome \cite{Farshid-04}.

\section{A stochastic model for nucleosome sliding}

\begin{figure}
\resizebox{1\columnwidth}{!}{
\includegraphics{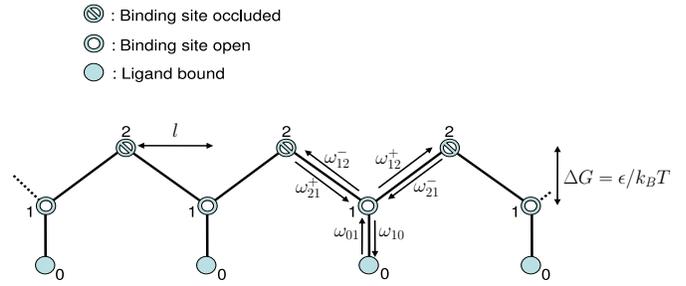}}
\caption{Schematic view of the three-state model for sliding of a
nucleosome on a periodic DNA sequence of height $\Delta G=
\epsilon k_BT$ and period $2l$. States 0, 1, and 2 denote states
where the ligand is bound, unbound with its binding site open and
unbound with its binding site closed, respectively.
$\omega_{ij}^{+(-)}$ denote the rates of going from state $i$ to
state $j$ to the right (left).}\label{fig_schem}
\end{figure}

In the absence of ligands, the sliding of the nucleosome along its
wrapped DNA can be described by a hopping model with a single
state or with two-states depending on the DNA sequence. The effect
of the sequence can be modelled as shown in Fig. \ref{fig_schem}
(without considering state $0$ in this figure). State $1$
represents the preferred binding sites for the DNA-histone complex
(the minimum of the sequence dependent potential), while state $2$
may be the high energy state of this potential. For uniform
sequences the height of the potential barrier is zero (the
energies of the two states are equal) that corresponds to a
single-state model. For a positioning sequence of DNA, states $1$
and $2$ are separated by 5 bp, i.e., half of the period of the
sequence denoted as $l$ in the figure. Similar two-state models
have been used to describe the motion of a linear molecular motor
on its track (such as a kinesin walking on a microtubule)
\cite{David-07,David-08}.

The presence of ligands affects the interaction of the nucleosome
with the DNA, and we describe this as an additional state $0$ that
branches off the state $1$ or $2$. When the ligand binds to the
DNA, we assume that no sliding can happen, as confirmed by
experiments \cite{Gottesfeld-02}. So the nucleosome waits until
the ligand detaches from the DNA, and this will reduce the
mobility of the nucleosome. The ligand may bind on the DNA either
in state $1$ and $2$ depending on the location of its substrate
that is a short DNA sequence. In this model we only consider the
case where the ligand can bind to the DNA when the nucleosome is
in state $1$. This is equivalent to $\Delta G
> 0$ in Fig. \ref{fig_schem}. In state 2
(5 bp, half the DNA pitch, away from state 1) the binding site is
then inaccessible since it faces the octamer surface. Similar
periodic sequential kinetic models with branching, jumping or
deaths have been studied in \cite{Kolomeisky-00}. Especially in
Ref. \cite{Farshid-04} the three state model in Fig.
\ref{fig_schem} has been examined in the same context as the
current paper for special cases of the transition rates.

To model the dynamics of the histone, consider the one-dimensional
lattice of Fig. \ref{fig_schem}. The nucleosome can hop to
neighboring sites on this lattice with some specific rates. Assume
that $l$ is the distance between sites $1$ and $2$ (which is half
the period), and let $p_i(n,t)$ be the probability for the
nucleosome to be in the states $i=0,1,2$, at position $x=nl$ and
at time $t$. The $p_i(n,t)$ satisfy the master equation
\begin{eqnarray}
\partial_t p_0(n,t)&=&\omega_{10}\,p_1(n,t)-\omega_{01}\,p_0(n,t)\label{Eq_P1}\\
\nonumber\\
\nonumber \partial_t
p_1(n,t)&=&\omega_{21}^+\,p_2(n-1,t)+\omega_{21}^-\,p_2(n+1,t) \label{Eq_P2}\\
&+& \omega_{01}\, p_0(n,t) -(\omega_{12}^++\omega_{12}^-+\omega_{10})\,p_1(n,t) \nonumber \\
\\
\nonumber\partial_t
p_2(n,t)&=&\omega_{12}^+\,p_1(n-1,t)+\omega_{12}^-\,p_1(n+1,t)\\
&-&(\omega_{21} ^++\omega_{21}^-)\,p_2(n,t),\label{Eq_P3}
\end{eqnarray}
where $\omega_{ij}^+$ ($\omega_{ij}^-$) represents the rate of
transition from state $i$ to neighboring state $j$ on the right
(left); and $\omega_{10}$ ($\omega_{01}$) represents the binding
(unbinding) rate of a ligand.

Let us introduce a vector $\vec{F}(\lambda,t)$, the components of
which are generating functions of the position for each state $i$.
These components are
$F_i(\lambda,t)=\displaystyle\sum_{n}{e^{-\lambda nl}p_i(n,t)}$.
The use of generating function reduces the space of configurations
that is infinite, to only 3 states due to the periodicity of the
system. It makes the calculations much more tractable. The master
equation now becomes:
\begin{equation}
\partial_t \vec{F}(\lambda,t)=M(\lambda) \vec{F}(\lambda,t),
\label{Dynamical eq}
\end{equation}
with
\begin{equation}
M(\lambda)= \left[
\begin{array}{ccc}
-\omega_{01} & \omega_{10} & 0 \\
&&\\
\omega_{01} & -\omega_{12}^+-\omega_{12}^--\omega_{10} & \omega_{21}^+\,e^{-\lambda l}+\omega_{21}^-\,e^{\lambda l} \\
&&\\
0 & \omega_{12}^+\,e^{-\lambda l}+\omega_{12}^-\,e^{\lambda l} & -\omega_{21}^+-\omega_{21}^-
\end{array}
\right].
\end{equation}
The solution of Eq. (\ref{Dynamical eq}) is
 \bb
 \vec{F}(\lambda,t) = e^{M(\lambda)t} \vec{F}(\lambda,0).
 \en
After calculating the eigenvalues of the matrix $M$, one can see that at long time
$t\rightarrow+\infty$ only the largest eigenvalue of $M$ that is
denoted by $\sigma_m(\lambda)$, contributes to $\displaystyle
\sum_{i}{F_i(\lambda,t)}=\left\langle e^{-\lambda nl}\right\rangle
\sim e^{\sigma_m(\lambda)t}$. Note that the normalization
condition for the probability implies that $\sigma_m(0)=0$. The
eigenvalue $\sigma_m(\lambda)$ contains all the long time
dynamical properties of the system, such as the velocity and the
diffusion constant \cite{David-07,David-08}, since
\begin{equation}
\overline{v}=-\left.\frac{d\sigma_m}{d\lambda}\right|_{\lambda=0},
\label{v}
\end{equation}
and
\begin{equation}
D=\left.\frac{1}{2}\frac{d^2\sigma_m}{d\lambda^2}\right|_{\lambda=0}.
\label{D}
\end{equation}

One can expand $\sigma_m(\lambda)$ near $\lambda=0$ as
$\sigma_m(\lambda)=-\overline{v}\lambda+D\lambda^2 +
O(\lambda^2)$. Using this expansion in the eigenvalue equation,
$\det \left[M-\sigma_m(\lambda)I_3 \right]=0$, the velocity and the diffusion
constant are derived as
\begin{eqnarray}
\overline{v}&=&\frac{2\left(\omega_{12}^+\omega_{21}^+-\omega_{12}^-\omega_{21}^-
\right)l}{S+\frac{\omega_{10}}{\omega_{01}}(\omega_{21}^++\omega_{21}^-)},
\label{explicit v}\\
\nonumber\\
D &=& 2l^2 \frac{\left(\omega_{12}^+\omega_{21}^++\omega_{12}^-\omega_{21}^-\right)
K + 8\omega_{12}^+ \omega_{12}^- \omega_{21}^+ \omega_{21}^- + \frac{\omega_{10}}
{\omega_{01}}J }{\left[S+\frac{\omega_{10}}{\omega_{01}}(\omega_{21}^++\omega_{21}^-)\right]^3}, \nonumber \\
\label{explicit D}
\end{eqnarray}
with
\begin{eqnarray}
S&=&\omega_{12}^++\omega_{12}^-+\omega_{21}^++\omega_{21}^-,\\
\nonumber \\
\nonumber K&=&{\omega_{12}^+}^2+{\omega_{12}^-}^2+{\omega_{21}^+}^2+{\omega_{21}
^- } ^2\\
&+&2
\left(\omega_{12}^+\omega_{12}^-+\omega_{12}^+\omega_{21}^-+\omega_{21}^+\omega_{21}^-+\omega_{21}^+\omega_{12}^-\right),\\
\nonumber\\
\nonumber
J&=&\left(\omega_{12}^+\omega_{21}^++\omega_{12}^-\omega_{21}
^-\right)\times\\
\nonumber &&\left(\omega_{21}^++\omega_{21}^-\right)
\left[2S+\frac{\omega_{10}}{\omega_{01}}\left(\omega_{21}^++\omega_{21}^-\right)\right]\\
&&-2\left(\omega_{12}^+\omega_{21}^+-\omega_{12}^-\omega_{21}^-\right)^2\left[1-\frac{\omega_{21}^++\omega_{21}^-}{\omega_{01}}\right].
\end{eqnarray}
It is worth to mention that in the ligand-free case the rate of
going from site 1 to 0 is zero, $\omega_{10}=0$, and the model
becomes equivalent to the two-state model that has been discussed
in \cite{David-07,David-08}.

\section{Kinetic rates in the absence of force}

We assume that the nucleosome sliding in the absence of any
external force is a passive process, which implies
that there is no preference between left and right direction.
Thus $\omega_{12}^+=\omega_{12}^-=\omega_{12}$ and
$\omega_{21}^+=\omega_{21}^-=\omega_{21}$. Let us introduce
$\kappa$ as the ratio of rates
\begin{equation}
\kappa \equiv \frac{\omega_{12}}{\omega_{21}}= e^{-\epsilon},
\end{equation}
where $\epsilon=\Delta G/k_B T$ is the energy difference between
the two states $1$ and $2$, and the above equality expresses the
detailed balance condition.

We consider the binding chemical reaction of the ligand:
$L+S\rightleftharpoons LS$, where $S$ is the substrate, i.e. the
nucleosomal DNA, $L$ is the ligand and $LS$ is the
ligand-substrate complex. From kinetics theory of first order
chemical reactions, one can write $\omega_{10}=k_+[L][S]$ and
$\omega_{01}=k_- [LS]$, such that the equilibrium constant of the
chemical reaction is $K_{eq}=k_+/k_-=[LS]_{eq}/\left(
[L]_{eq}[S]_{eq} \right)$, in terms of equilibrium concentration
of ligand $[L]_{eq}$, substrate $[S]_{eq}$ and  complex
$[LS]_{eq}$. Therefore the ratio
 \bb
 \eta \equiv \frac{\omega_{10}}{\omega_{01}}=K_{eq} \frac{[L][S] }{[LS]} =
 \frac{[LS]_{eq}}{[L]_{eq} [S]_{eq}} \frac{[L][S] }{[LS]}, \label{eq_eta}
 \en
 quantifies the deviation away from equilibrium.
In general, the substrate is in excess so that $[S]$ and $[S]_{eq}$
are both large, and also $[S] \simeq
[S]_{eq}$. Furthermore, if one assumes that $[LS] \simeq [LS]_{eq}$,
then $\eta \simeq [L]/[L]_{eq}$.

From Eq.~(\ref{explicit v}), one finds that $\overline{v}=0$ as
expected and the diffusion constant is:
\begin{equation}
D=\frac{2\;\omega_{12}\;l^2}{1+\eta+\kappa},\label{eq_D}
\end{equation}
where $\omega_{12}$ can be calculated by Kramers rate theory in
the limit $\epsilon \gg 1$.

The sequence dependent potential can be modelled as a periodic
function $\beta U(x)=\frac{\epsilon}{2}\, \cos\left(\frac{\pi
x}{l}\right)$ with $\beta=1/k_BT$ \cite{Kulic-03}. The time needed
for the nucleosome to go from one minimum of this potential to the
neighboring maximum, $\tau$, can be determined using the Kramers
rate theory :
\begin{equation}\frac{1}{\tau}= \frac{D_0}{2\pi k_BT}\sqrt{\left|U''(1)U''(2)\right|}\;
e^{-\epsilon}=\frac{\pi\left|\epsilon\right|}{4l^2}D_0\;e^{-\epsilon},\label{tau_eq}
\end{equation}
where $D_0$ is the diffusion constant of the nucleosome without
any potential barrier or ligand, $1$ and $2$ refer to the minimum
and the maximum of the potential $U(x)$. From this we find
$\omega_{12}$ for such a strong positioning sequence:
\begin{eqnarray}
\omega_{12}=\frac{1}{2\tau}
=\frac{\pi\epsilon}{8l^2}D_0\;e^{-\epsilon} \qquad {\rm for \; \; } \epsilon>0 ,\label{eq_w}
\end{eqnarray}
with the factor $\frac{1}{2}$ being the probability to go through
the barrier $2$ from either direction \cite{Hanngi-1990}. It is
worth to mention that in the limit $\epsilon \gg 1$ with
($\epsilon>0$, $\eta \geq 0$) when there is a sequence dependent
potential for the nucleosome,
 one can find that:
\begin{equation}
D=\frac{\pi\epsilon\, e^{-\epsilon}}{4(1+\eta)}D_0. \label{eq_D2}
\end{equation}
In the case of random sequence of DNA, i.e in the absence of a
sequence dependent potential,
$\epsilon = 0$ and then $\omega_{12}$ is simply equal to
\begin{equation}
\omega_{12}=\frac{D_0}{l^2},\label{eq_omega_0}
\end{equation}
and the diffusion constant can be found as
\begin{equation}
D=\frac{2D_0}{2+\eta}.
\end{equation}
Putting in numbers, using realistic parameter values \cite{Kulic-03} $D_0\simeq600\,bp^2/s$, $\epsilon=9$,
$l=5\,bp$, $\eta=0 $ in the absence of ligand, and $\eta\simeq100 $ in the presence of ligands our model
predicts the time needed for a nucleosome to diffuse on a $70$ bp DNA, to be $78$ minutes without ligands and
$131$ hours with the ligands which is consistent with the experimental observations \cite{Gottesfeld-02}. For a
random sequence ($\epsilon=0$), in the presence ($\eta\simeq100 $) and the absence of ligands ($\eta \simeq 0
$), the characteristic time for 70 bp diffusion are 3.5 min and 4 sec, respectively.

In the absence of ligands and for arbitrary sign of $\epsilon$,
Eq.~(\ref{eq_D2}) becomes
\begin{equation}
D=\frac{\pi|\epsilon| \, e^{-|\epsilon |}}{4}D_0. \label{eq D3}
\end{equation}
The above expression that has been derived
with a discrete stochastic model, can be also derived using a
continuous description in the limit $\epsilon \gg 1$. This can be
done by considering a particle diffusing in the periodic
potential $U(x)$, for which the diffusion coefficient is
\cite{Hanngi-1990}
\begin{equation} \label{D periodic}
D=D_0 \langle e^{\beta U(x)} \rangle ^{-1} \langle e^{-\beta U(x)}
\rangle ^{-1}.
\end{equation}
Indeed $\langle e^{\beta U(x)} \rangle$ is a Bessel function, the
asymptotic form of which is $2 e^{\epsilon/2}/\sqrt{\pi \epsilon}$
for $\epsilon \gg 1$. From this Eq.~(\ref{eq D3}) is recovered
using Eq.~(\ref{D periodic}).

\subsection{Transcription-induced sliding}

Up to now we considered thermally induced, undirected nucleosome
sliding. Here we discuss the case when a force is applied to the
nucleosome. As discussed in the introduction this situation might
occur when an RNA polymerase encounters a nucleosome during
transcription. A more microscopic model of such an encounter is
presented in Ref. \cite{Laleh-08}. Note that we will assume that
the polymerase does not unpeel the nucleosome, a case considered
recently by T. Chou \cite{Chou-07}. Rapid progress in the field of
micromanipulation experiments let us expect that there will be
also soon data available where forces are directly applied to
nucleosomes.

A force $F$ exerted on the nucleosome introduces a bias in the
transition rates:
\begin{eqnarray}
\omega_{12}^+&=&\omega\;e^{-\epsilon}\;e^{\theta_1^+f},\label{Eq_omega12+}\\
\omega_{21}^-&=&\omega\;e^{-\theta_2^-f},\label{Eq_omega21-}\\
\omega_{12}^-&=&\omega\;e^{-\epsilon}\;e^{-\theta_1^-f},\label{Eq_omega12-}\\
\omega_{21}^+&=&\omega\;e^{\theta_2^+f},\label{Eq_omega21+}
\end{eqnarray}
where $\theta_i^{\pm}$ are the load distribution factors
\cite{Kolomeisky-00}, and $f\equiv Fl/k_BT$. Using detailed
balance condition, one has
$\theta_1^++\theta_1^-+\theta_2^++\theta_2^-=2$.

Putting these rates in Eq. (\ref{explicit v}), the velocity
of the nucleosome can be found as
\begin{equation}
v=2\omega l \;e^{-\epsilon}\frac{e^{(\theta_1^++\theta_2^+)f}-e^{-(\theta_1^-+\theta_2^-)f}}
{e^{-\epsilon}\left(e^{\theta_1^+f}+e^{-\theta_1^-f}\right)+
(1+\eta)\left(e^{\theta_2^+f}+e^{-\theta_2^-f}\right)}\label{eq_vF}
\end{equation}

The mobility of nucleosome is defined as
\begin{equation}
\mu \equiv \frac{1}{k_BT} \left. \frac{dv}{df}\right|_{f=0}, \label{eq:def_mu}
\end{equation}
which gives
\begin{equation}
\mu=\frac{1}{k_BT}\frac{2\omega \kappa}{1+\eta+\kappa}l^2
\end{equation}
Comparing this expression with Eq. (\ref{eq_D}) one finds the
Einstein relation $\mu= D/(k_BT)$ verified.

\section{Results}

\begin{figure}
\resizebox{0.9\columnwidth}{!}{
\includegraphics{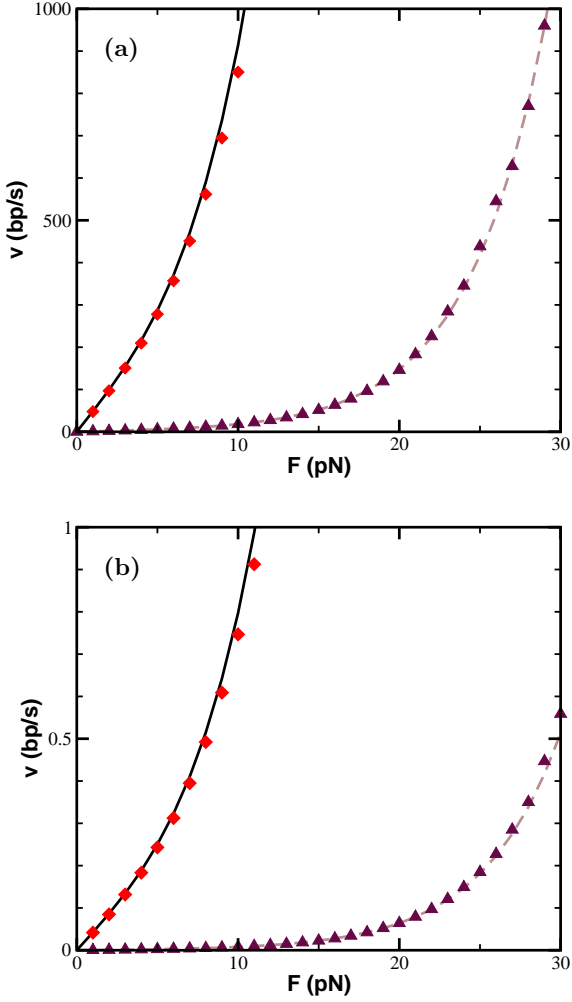}}
\caption{The velocity of the nucleosome versus the external force
exerted on the DNA for (a) a random sequence of DNA ($\epsilon=0$)
and (b) a positioning sequence with $\epsilon=9$ in two cases: the
diamonds are simulation data for $\eta=0$ while the black solid
line is plotted using theory, Eq. (\ref{eq_vF}), the triangles are
simulation data for $\eta=100$ while the brown dashed line is
plotted using theory, again Eq. (\ref{eq_vF}).}\label{fig_vF}
\end{figure}

There are three physical quantities that affect the behavior of
the system: the external force $F$, the sequence dependent part of
the potential measured by
$\epsilon$ and the ligand concentration $[L]$ that enters into
$\eta$ through $\eta \simeq [L]/[L]_{eq}$. The physical behavior
of the system is characterized by the velocity of the nucleosome
repositioning along the DNA and its diffusive behavior. In this
section, the results of the analytical approach described in the
previous sections and of a computer simulation that is discussed
in Appendix A, are presented.

We first consider the effect of an external force on the velocity
of the nucleosome repositioning along the DNA. In Fig.
\ref{fig_vF} we plot the nucleosome velocity $v$ versus the
applied force $F$ for the two limiting cases $\eta=0$ and
$\eta=100$ of the experiments \cite{Gottesfeld-02}, on both random
DNA and on a positioning sequence. As expected, the velocity of
the nucleosome increases with the external force and there is a
good agreement between the simulation results and the analytical
approach. At zero force, the nucleosome shows purely diffusive
behavior and there is no net velocity, cf. Fig. \ref{fig_vF} at
$F=0$. As soon as a force is applied, there is a bias in the
transition rates and the nucleosome attains a drift velocity in
the direction of the applied force. A positioning sequence of DNA
leads to an effective potential barrier on the corkscrew path of
the nucleosome \cite{Kulic-03} leading to a drift that is
significantly smaller than on random DNA. The aforementioned
behaviors are seen in Fig. \ref{fig_vF}.

\begin{figure}
\resizebox{0.9\columnwidth}{!}{
\includegraphics{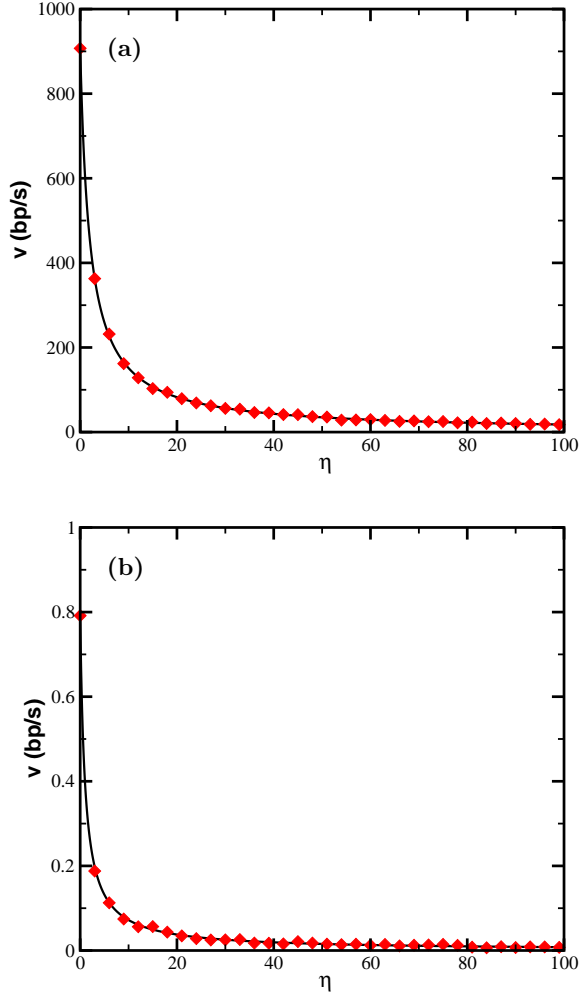}}
\caption{Velocity $v$ of the nucleosome versus $\eta$ in the
presence of an external force $F=10\,pN$ on (a) a random basepair
sequence ($\epsilon=0$) and (b) a positioning element
($\epsilon=9$). The red diamonds are the simulation results while
the line is plotted using the analytical approach, Eq.
\ref{eq_vF}.} \label{fig_vL}
\end{figure}

Next we study how the ligands influence the sliding velocity. In
Fig. \ref{fig_vF} we see that for a given finite force the
nucleosome velocity for the case $\eta=100$ is much smaller than
in the absence of ligands, $\eta=0$. The effect of the ligand
concentration on the drift velocity of the nucleosome for a
typical external force, $F=10\,pN$, is shown in Fig. \ref{fig_vL}.
As expected ligands block corkscrew sliding, lowering the overall
drift velocity. For typical experimental numbers
\cite{Gottesfeld-02} already a small concentration of ligands
significantly lowers the drift, cf. Fig. \ref{fig_vL}.

\begin{figure}
\resizebox{0.9\columnwidth}{!}{
 \includegraphics{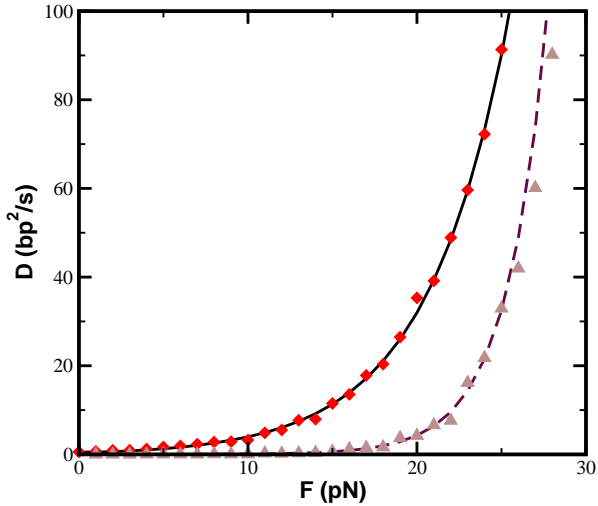}}
 \caption{Diffusion constant $D$ of the nucleosome versus external
 force $F$ on a positioning sequence ($\epsilon = 9$). The solid
 black line (red diamonds) and the brown dashed line
 (light brown triangles) are the theoretical (simulation) results
 for $\eta = 0$ and $\eta = 100$. }\label{fig_DF9}
 \end{figure}

Another parameter that provides information about the system is
the diffusion constant of the nucleosome, $D$. The behavior of $D$
versus both the external force and the ligand concentration has
been checked on for random DNA and a positioning sequence. In Fig.
\ref{fig_DF9} it can be seen that $D$ increases with $F$. Using a
simple two state model can help us to understand the behavior of
the diffusion constant in terms of force $F$. This simple model has been
explained in Appendix B. We see that in the presence of external force,
the diffusion constant of the particle becomes larger when the force
is increased.

\begin{figure}
\resizebox{0.87\columnwidth}{!}{
\includegraphics{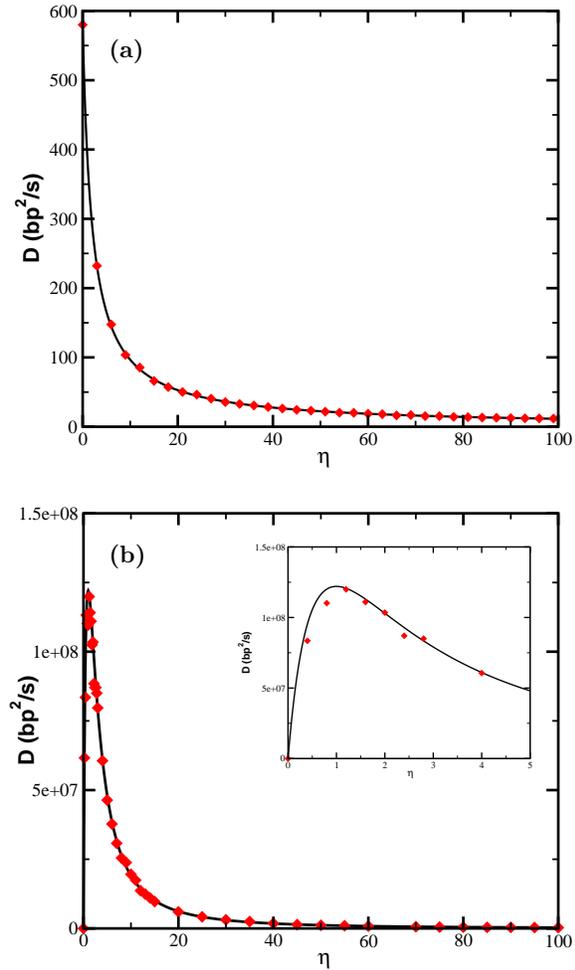}}
\vspace{.1cm}\caption{Nucleosome diffusion constant $D$ versus
$\eta$ for random DNA ($\epsilon=0$) for two different forces: (a)
$F=0\,pN$ and (b) $F=10\,pN$ (the inset gives a zoomed view
showing the non-monotonous behavior for small $\eta$-values). The
red diamonds are the simulation results while the lines use the
analytical approach, Eq. \ref{explicit D}. }\label{fig_DL0}
\end{figure}

\begin{figure}
\resizebox{0.87\columnwidth}{!}{
\includegraphics{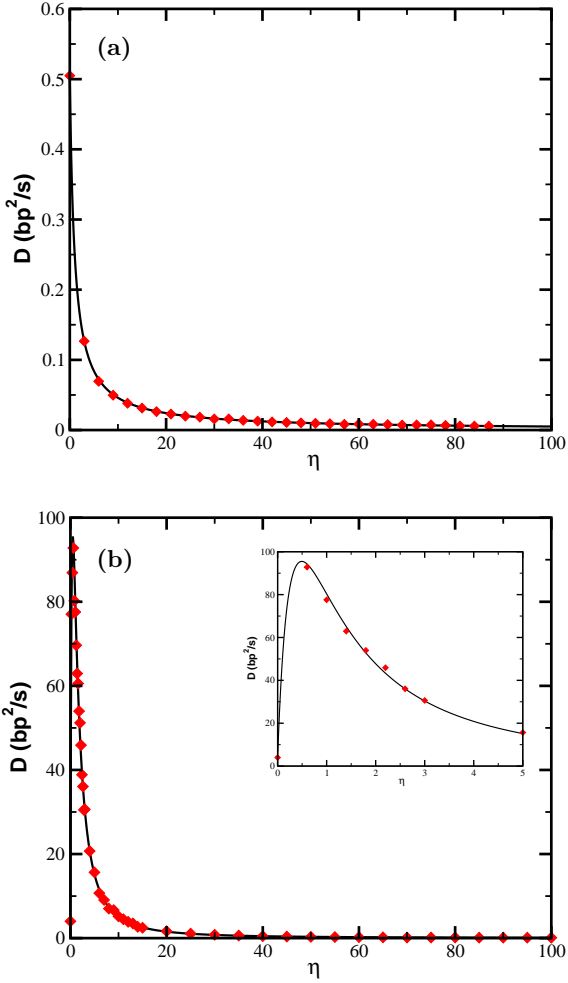}}
\caption{The diffusion constant of the nucleosome versus $\eta$ in
the case of a positioning sequence with $\epsilon=9$ for two
different forces: (a) $F=0\,pN$ and (b) $F=10\,pN$. The inset is
the zoomed-in plot showing the behavior of the diffusion constant
for small values of $\eta$. The red diamonds are the simulation
results while the lines are plotted using the analytical
approach.}\label{fig_DL9}
\end{figure}

We present the behavior of the diffusion constant versus $\eta$,
for two different forces, $F=0 pN$ and $F=10\, pN$, in Figs.
\ref{fig_DL0} and \ref{fig_DL9}. Naively one would expect that at
a fixed external force the diffusion constant decreases with
$\eta$ since an increase of the ligand concentration leads to a
higher probability to have a ligand bound that then suppresses
diffusion. For zero force the diffusion constant does indeed
follow this expectation, cf. Figs. \ref{fig_DL0}(a) and
\ref{fig_DL9}(a). Interestingly, in the presence of a nonzero
external force the behavior of the diffusion constant versus
$\eta$ differs dramatically from this expectation. For $\eta
\lesssim 1$, $D$ increases with $\eta$, and then decreases as
$\eta$ goes to infinity (Figs. \ref{fig_DL0}(b) and
\ref{fig_DL9}(b)). For random (positioning) DNA the maximal value
of the diffusion constant is five (two) orders of magnitude larger
than the value in the absence of ligands.

\begin{figure}
\resizebox{0.9\columnwidth}{!}{
\includegraphics{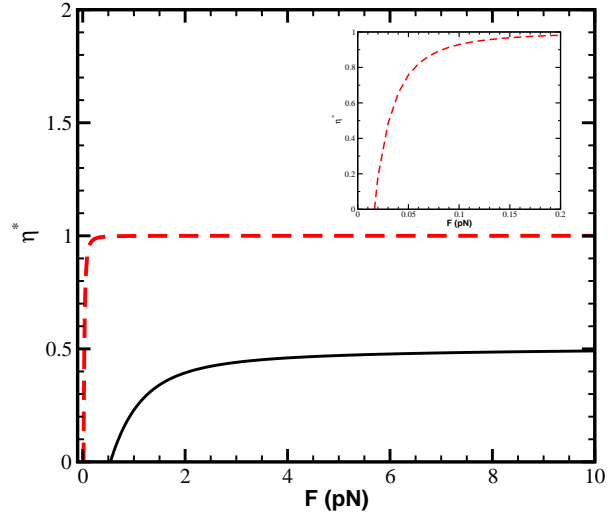}}
\caption{$\eta^*$ versus $F$ for the case $\theta_i^\pm=1/2$ for
random DNA (dashed line) and a nucleosome positioning element
(solid line). The lines are drawn using Eq. \ref{EQ_eta*}.
}\label{fig_eta}
\end{figure}

For $\eta \ll 1$ the diffusion constant in 3-state model behaves
as the 2-state model with a correction like
$D_{\rm{3-state}}=D_{\rm{2-state}} +O(\eta)$. For large values of
$\eta$, as can be seen in Eq. (\ref{explicit D}), the diffusion
constant changes as $D\propto 1/\eta$ and the diffusion constant
decreases as $\eta$ increases. The $\eta$ at which $D$ attains its
maximum $\eta^*$ can be calculated from Eq. (\ref{explicit D}) and
is shown in Fig. \ref{fig_eta} as a function of $F$ for two cases
of random and positioning DNA. As can be seen in this figure, at
not very small forces, $\eta^*$ is equal to $1$ for random DNA and
$0.5$ for the positioning sequence. Since at large forces the
positive rates $\omega^+$ dominate, the diffusion constant, Eq.
(\ref{explicit D}), simplifies to
\begin{eqnarray}
 D=2l^2\frac{\eta(\omega_{12}^+\omega_{21}
^+)^2\omega_{21}^+}{\omega_{01}\left[\omega_{21}
^+(\eta+1)+\omega_{12}^+\right ]^3}.\label{EQ_D-bigF}
\end{eqnarray}
Setting the derivative of $D$ with respect to $\eta$ equal to
zero, one obtains
\begin{eqnarray}
 \eta^*&=&\frac{\omega_{12}^++\omega_{21}^+}{2\omega_{21}^+} \nonumber \\
&=&\frac{1+e^{-\epsilon+(\theta_1^+-\theta_2^+)f}}{2},
\label{EQ_eta*}
\end{eqnarray}
where we have used Eqs. (\ref{Eq_omega12+})--(\ref{Eq_omega21+}).
For the case $\theta_i^\pm=1/2$ the force dependent term drops out
and we find $\eta^*=(1+e^{-\epsilon})/2$ and especially
$\eta^*=1$ for random DNA. We shall come back to this point in the
discussion section. In Fig. \ref{fig_eta}, the behavior of
$\eta^*$ is plotted versus the external force in the case of
$\theta_i^\pm=1/2$.

Let us now discuss the behavior of the diffusion constant versus
$\eta$ for small values of $\eta$, i.e. for $\eta \ll 1$. Through
an expansion of the exact expression we obtain
\begin{eqnarray}
D = A_0(f) + A_1(f)\eta + O(\eta^2),\label{Eq_D_expand}
\end{eqnarray}
where $A_0$ and $A_1$ are functions of $f$ that are given in
Appendix C. It is convenient to expand $A_1(f)$ as a function of
$f$:
\begin{eqnarray}
A_1(f) = \left.\frac{\partial D}{\partial \eta}\right |_{\eta=0} =
\alpha(\beta_0+ \beta_1 f^2) + O(f^4),\label{Eq_D'}
\end{eqnarray}
where $\alpha$, $\beta_0$, and $\beta_1$ are functions defined in
Appendix C. Since $\alpha >0$ and $\beta_0<0$, we have $\partial
D/\partial \eta<0$ for both random and positioning sequence of DNA
at $f=0$ that can indeed be seen in the plots of diffusion
constant $D$ versus $\eta$ shown in Figs. \ref{fig_DL0}(a) and
\ref{fig_DL9}(a). When $\beta_1$ becomes positive, there is a
threshold in force that is $f_T=\sqrt{-\beta_0/\beta_1}$ such that
for $f>f_T$ the derivative of $D$ with respect to $\eta$ is
positive for small $\eta$-values. From the expression of the
coefficients derived in Appendix C, the value of $f_T$ is the
following function of $\epsilon, \omega$ and $\omega_{01}$:
\begin{eqnarray}
f_T = \frac{\sqrt{2} \left( 1 + e^{-\epsilon} \right)}
{\left[ -3 + e^{-\epsilon} \left( 5- 6 e^{-\epsilon}\right) +
4 e^{-\epsilon} \left( 1 + e^{-\epsilon}\right) \frac{\omega}{\omega_{01}} \right]^{1/2}}.
\label{eq:fT}
\end{eqnarray}
Putting in numbers, $\omega_{01} = 0.001 \; s^{-1}$ and $\omega =
D_0/l^2 \simeq 24 \; s^{-1}$, we find $f_T \simeq 0.48$ that is
equivalent to $F \simeq 0.016 \; pN$ for a random sequence of DNA.
In the case of positioning sequence of DNA the rate of $\omega$
changes to $\omega = \pi \epsilon D_0 / (8 l^2) \simeq 85 \;
s^{-1}$ and we find $f_T \simeq 0.23$ that is equivalent to $F
\simeq 0.55 \; pN$. In Fig. \ref{fig_eta} the behavior of $\eta^*$
as a function of $F$ is shown for both random and positioning
sequence of DNA. We note that the force value for which $\eta^* =
0$ corresponds to the threshold force given by Eq. (\ref{eq:fT}).

How can the surprising non-monotonous behavior of $D$ as a
function of $\eta$ of a
nucleosome driven by the application of a force be explained,
especially the strongly enhanced fluctuations around $\eta^*$ with
$D(\eta^*) \gg D(\eta=0)$? Obviously the fluctuations of position
of this driven nucleosome in the presence of ligands are of
different origin than the ones in the absence of ligands. For
sufficiently large forces the nucleosome mostly steps in the
direction of the force or -- if the nucleosome is in state 1 -- a
ligand might bind. The latter event stops the drifting nucleosome
for a while and is thus a source of
fluctuations of completely different origin, because this
introduces some waiting time before hopping from states 1. The
higher the concentration of ligands, i.e. the higher the value of
$\eta$, the more often these events occurs, increasing their
contribution to the overall fluctuation of the nucleosome, which
are measured by the diffusion coefficient. This is the case up to
a critical value of $\eta$, named $\eta^*$, which is force
dependent. Further addition of ligands populates the state 0 so
much that the nucleosomes gets frequently stuck, decreasing its
diffusion constant.

\section{Discussion}

In our model the external force changes the local rates from sites
$1$ to $2$ and vise versa. This is because the force produces
internal stress on the nucleosomal DNA that introduces a bias in
the dynamics. This assumption can be verified by considering the
microscopic details of the interaction of the nucleosome with the
DNA that will be presented in a forthcoming publication
\cite{Laleh-08}.

In the above treatment we assumed that both binding and unbinding
rates of ligands to the DNA do not depend on the external force.
The typical length of a ligand site is about 6-7 bps and the DNA
length between two adjacent binding sites is 10 bps. Even at the
highest forces considered  here (30 pN) the probability of having
a defect in a ligand binding region is close to zero
\cite{Laleh-08} so that we expect that its influence on the ligand
rates can be neglected.

It is also important to point out that the load distribution
factors $\theta_i^{\pm}$ have an effect on the diffusion constant. We
assumed that all $\theta_i^{\pm}$ are equal to $\frac{1}{2}$. By
changing the values of $\theta_i^\pm$ the overall behavior of all
plots does not change, although the precise values of the
diffusion coefficient and even the curvature of the plots can be
affected. For instance, the value of $\eta^*$ at which $D$ is
maximized depends on the values of $\theta_i^\pm$ as can be seen
from Eq. (\ref{EQ_eta*}). The microscopic details of the
interaction between the DNA and the nucleosome determine the
values of the distribution factors. For a random sequence there is
no reason to have different rates for backward and forward steps
of the nucleosome along the DNA. Therefore, one finds that for
large values of the external force, $\eta^*$ converges to $1$. For
a DNA positioning sequence, the values of $\theta_i^\pm$ could in
principle depend on the sequence. The exact value of these
coefficients can only be determined from experimental data or from
a more detailed modelling of the transition state. We have
arbitrarily chosen them to be $1/2$ for the plots. Note that if
$\theta_1^+<\theta_2^+$ then $\eta^*|_{F \rightarrow \infty } =
0.5$, the same value as in our case $\theta_1^+=\theta_2^+$, while
for $\theta_1^+>\theta_2^+$, $\eta^*$ increases as $F$ is
increased and one has $\eta^*|_{F \rightarrow \infty} \rightarrow
\infty$.

\section{Acknowledgments}

We are thankful to K. Mallick and H. Fazli for valuable discussions.
D. L. acknowledges support from the the Indo-French Center CEFIPRA (grant 3504-2)
and F. M. acknowledges support from CNRS and the hospitality of
Laboratoire de Physico-Chimie Th\'eorique, UMR 7083, ESPCI in Paris
where this work was initiated.


\section{Appendix}

\appendix

\section{The algorithm of the simulation}

The 3-state model presented in this paper is simulated using a
``Random Selection Method'' \cite{random alg}. It is defined in
terms of the transition rates $\omega_{ij}^\pm$ that give the
probabilities per unit time for going from state $i$ to state $j$
in the plus/minus direction. If the system is at time $t$ in the
state $i$, a transition to the neighboring state $j$ happens at
time $t\,+\,\Delta t$ with the finite probability
$P_{ij}^\pm=\Delta t \, \omega_{ij}^\pm$.

For each step, a random number $0\leq \xi < 1$ is drawn.
Depending on its value and the state of the system a
decision is taken:

\begin{eqnarray}
\nonumber\rm{state} : 0 &\rightarrow& 1 ~ \quad ~\rm{if} ~ 0
\le \xi< P_{01},\\
\nonumber\rm{state} : 0 &\rightarrow& 0 ~\quad ~
\rm{otherwise}\\
\nonumber\\
\nonumber\rm{state} : 1 &\rightarrow& 0 ~ \quad ~~ \rm{if} ~
0 \le \xi < P_{10},\\
\nonumber\rm{state} : 1 &\rightarrow& 2^+ ~ \quad \rm{if} ~
P_{10}
\le \xi < P_{10}+P^+_{12},\\
\nonumber\rm{state} : 1 &\rightarrow& 2^- ~ \quad \rm{if} ~
P_{10}+P_{12}^+ \le \xi < P_{10}+P_{12}^++P_{12}^-,\\
\nonumber\rm{state} : 1 &\rightarrow& 1 ~ \quad ~~
\rm{otherwise}\\
\nonumber\\
\nonumber\rm{state} : 2 &\rightarrow& 1^+ ~ \quad \rm{if} ~
0 \le
\xi < P_{21}^+,\\
\nonumber\rm{state} : 2 &\rightarrow& 1^- ~ \quad \rm{if} ~
P_{21}^+ \le \xi < P_{21}^++P_{21}^-,\\
\nonumber\rm{state} : 2 &\rightarrow& 2 ~ \quad ~ ~
\rm{otherwise}.
\end{eqnarray}

For the next time step, from $t+\Delta t$ to $t+2\Delta t$, the
procedure is repeated again. The time step $\Delta t$ is chosen
small enough, such that for each step the condition $\sum
\omega_{ij}^\pm \Delta t < 1 $ is satisfied, where the sum is
taken over the probabilities of all possible transitions from
state $i$.

This algorithm is similar to the one of Gillespie
\cite{Gillespie-77}, except for the fact that we use here a
constant time step whereas for the Gillespie algorithm the time
step is a random variable. Both algorithms converge to the same
steady state albeit after different times as we also checked for
our model. The steady state probabilities for the three states are
obtained by setting the time derivatives of the probabilities in
Eqs. (\ref{Eq_P1})--(\ref{Eq_P3}) to zero:
\begin{eqnarray}
p_0&=&\frac{\eta}{1+\eta+e^{-\epsilon}},\\
p_1&=&\frac{1}{1+\eta+e^{-\epsilon}},\\
p_2&=&\frac{e^{-\epsilon}}{1+\eta+e^{-\epsilon}},
\end{eqnarray}
a result that has been previously obtained in Ref.
\cite{Farshid-04}. We let the simulation run for a long time (from
$t_0$ to $t_N$ with $t_i=i\Delta t$, $N\gg 1$) to be sure that the
system has reached equilibrium. Then averaged over $M$ ensembles
with $M\gg 1$, the mean velocity and the diffusion constant is
determined by
\begin{eqnarray}
v=\frac{\sum_{r=1}^M(X_r(t_N)-X_r(t_1))}{N\Delta t\times M},
\end{eqnarray}
and $D$ is determined as the slope of the plot of
$\sum_{r=1}^M X_r^2/M - (\sum_{r=1}^M X_r/M)^2$ versus
$2\sum_{r=1}^M t_r/M$.

The time steps used for the simulation are less than $0.001$,
depending on the simulated case. The time goes to $10^5\,s$, and
the number of ensembles are $M=2000$. The used parameters for the
simulation are $\theta_i^{\pm}=1/2, ~ i=1,2$ and $\omega$ is
determined from to Eq. (\ref{eq_w}) and Eq. (\ref{eq_omega_0}) for
the positioning and random DNA sequence respectively. Also using
the experimental data, the typical time needed for a ligand to
unbound from the DNA is some minutes and
$\omega_{01}=0.001\,s^{-1}$.

\section{The behavior of the diffusion constant versus the
external force in a simple two state model}

\begin{figure}
\resizebox{0.9\columnwidth}{!}{
 \includegraphics{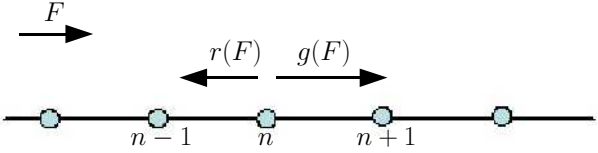}}
 \caption{The two state problem used to explain the behavior of the diffusion
 constant versus the applied force. The external force, $F$, changes the
 jumping  rates of the particle, which are shown by $g(F)$ and
$r(F)$.}\label{fig_schem_app}
 \end{figure}

To check the effect of force on the diffusion constant, let us assume a simple
two state model in which the external force changes the rates as shown in
Fig. \ref{fig_schem_app}. The position of the particle in the mentioned
lattice model is denoted by $n$. By definition, the diffusion constant is written as
\begin{equation}
D \equiv \frac{1}{2} \lim_{t\rightarrow\infty}\frac{\partial}{\partial
t}\left[\langle n^2\rangle - \langle n \rangle^2\right],
\end{equation}
where $\langle A \rangle$ denotes the average of quantity $A$ that is given by
$\langle A \rangle = \sum_n A_n P_n$. $P_n$ is the probability for the particle
to be in the position $n$.

The master equation governing this system can be written as
\begin{eqnarray}
 \frac{d\,P_n}{dt} = r(F) P_{n+1} + g(F) P_{n-1}
- \left[ r(F) + g(F) \right]P_{n},
\end{eqnarray}
where the force is denoted by $F$, $a$ is the jump length, and $g(F)$ and
$r(F)$ are the force-dependent rates for going to the right and left,
respectively. Depending on the direction the force is exerted on the system,
one of these rates increases and one decreases. Here, we have assumed
that the force pushes the system to the right, so $g(F)$ increases
with $F$ while $r(F)$ decreases with $F$. Then a simple calculation leads to:
\begin{eqnarray}
\frac{\partial}{\partial t}\langle n^2\rangle &=&
2 \left[g(F)-r(F)\right]\langle n\rangle + g(F)+r(F),\label{EQ_n2}\\
\frac{\partial}{\partial t}\langle n\rangle^2
&=& 2\left[g(F)-r(F)\right]\langle n\rangle,\label{EQ_n}
\end{eqnarray}
where we have used:
\begin{eqnarray}
 \frac{\partial}{\partial t}\langle A\rangle& = &
\sum_{n}A_n \frac{\partial}{\partial t}P_n,\\
\frac{\partial}{\partial t}\langle A\rangle^2 &=&
2\langle A \rangle \frac{\partial}{\partial t}\langle A_n\rangle,
\end{eqnarray}
and
\begin{eqnarray}
\sum_n P_n = 1.
\end{eqnarray}

Using Eqs. (\ref{EQ_n2}) and (\ref{EQ_n}), one finds the diffusion constant as
following:
\begin{equation}
 D=\frac{g(F)+r(F)}{2}. \label{eq:App-D-force}
\end{equation}
If the rates in the absence of the external force are denoted by $\omega$,
the external force, $F$, changes the jumping rates of the particle to
$g(F) = \omega \exp\left[{+F a /(2k_BT)}\right]$ and $r(F) = \omega
\exp\left[{-F a /(2 k_BT)}\right]$. Using Eq. (\ref{eq:App-D-force})
and the mentioned rates in the presence of
external force, the diffusion constant is found as
\begin{equation}
D = \frac{\omega}{2} \left(
e^{+F a /2 k_B T} + e^{- F a /2 k_B T}
\right).
\end{equation}
This explains the behavior of our results for the diffusion constant
mentioned in the text.

\section{The explicit forms of the auxiliary functions}

In this appendix we give the explicit form of the constants used
in the Eqs. (\ref{Eq_D_expand}) and (\ref{Eq_D'}). First we expand
$D$ for $\eta \rightarrow 0$:
\begin{eqnarray}
\nonumber D &=& \frac{ B_0 + B_1 \eta + B_3 \eta^2}{[S + B_2\eta]^3}\\
 &=& \frac{B_0}{S^3}+ \frac{B_1 S - 3 B_0 B_2 }{S^4}\eta +O(\eta^2),
\end{eqnarray}
where
\begin{eqnarray}
\nonumber B_0 &=& 2l^2 \left[ K(\omega_{12}^+\omega_{21}^+ +
\omega_{12}^-\omega_{21}^-) +
8\omega_{12}^+\omega_{21}^+\omega_{12}^-\omega_{21}^-
\right],\\
\nonumber B_1 &=& 2l^2 [2(\omega_{12}^+\omega_{21}^+ +
\omega_{12}^-\omega_{21}^-) (\omega_{21}^++\omega_{21}^-) S \\
\nonumber && -2(\omega_{12}^+\omega_{21}^+ - \omega_{12}^-\omega_{21}^-)^2 (1-
\frac{\omega_{21}^- + \omega_{21}^+}{\omega_{01}})] ,\\
\nonumber B_2 &=& \omega_{21}^+ + \omega_{21}^-,\\
\nonumber B_3 &=& 2l^2 \left[(\omega_{21}^- +
\omega_{21}^-)^2(\omega_{12}^+\omega_{21}^+ +
\omega_{12}^-\omega_{21}^-)\right].
\end{eqnarray}
From this follow the expansion coefficients in Eq.
(\ref{Eq_D_expand}):
\begin{eqnarray}
A_0(f)&=&\frac{B_0}{S^3},\\
A_1(f)&=&\frac{B_1 S - 3B_0 B_2 }{S^4}.
\end{eqnarray}

Finally we provide here the behavior of $A_1(f)$ for small forces.
Using Eqs. (\ref{Eq_omega12+})--(\ref{Eq_omega21+}) with
$\theta_i^\pm = 0.5$, the expansion of $B_i(f)$ for small forces
can be written as
\begin{eqnarray}
\nonumber S &=&2\omega(e^{-\epsilon}+1)(1+\frac{f^2}{4}),\\
\nonumber B_0 &=& 8 \omega^4e^{-\epsilon}(e^{-\epsilon}+1)^2 \\
&+&
4\omega^4e^{-\epsilon} \left[2+2
e^{-2\epsilon}+(e^{-\epsilon}+1)^2\right] f^2, \nonumber \\
\nonumber \frac{B_1}{2l^2} &=&
16\omega^4e^{-\epsilon}(e^{-\epsilon}+1) \\
&+& \left[16\omega^4e^{-2\epsilon}
\left(1+\frac{\omega}{ \omega_ { 01 } } + 24
\right)\right]f^2, \nonumber\\
\nonumber \frac{B_2}{2l^2} &=&2\omega(1+\frac{f^2}{4}).
\end{eqnarray}
Consequently,
\begin{eqnarray}
A_1(f) &=& \frac{16 l^2 \omega^5}{S^4} e^{-\epsilon} \left\{ - 2 \left( 1 + e^{-\epsilon}\right)^2 \right.
\nonumber \\
 &+& \left. \left[-3 + e^{-\epsilon } \left( 5 - 6 e^{-\epsilon} \right)
 + 4 e^{-\epsilon} \left( 1 + e^{-\epsilon} \right) \frac{\omega}{\omega_{01}} \right] \right\}, \nonumber \\
 \label{Eq_f_C}
\end{eqnarray}

Using these expansions and Eq. (\ref{Eq_D'}), one can write:
\begin{eqnarray}
\alpha&=& \frac{16 l^2 \omega^5}{S^4} e^{-\epsilon},\\
\beta_0&=&-2(1+ e^{-\epsilon})^2, \\
\beta_1 &=& -3 + e^{-\epsilon } \left( 5 - 6 e^{-\epsilon} \right)
 + 4 e^{-\epsilon} \left( 1 + e^{-\epsilon} \right) \frac{\omega}{\omega_{01}}.
\end{eqnarray}

\end{document}